\documentclass[twocolumn,runningheads,natbib]{svjour2}
\smartqed  
\usepackage{graphicx}
\usepackage{natbib}
\usepackage{psfig}

\newcommand{\ecs}{erg\,cm$^{-2}$\,s$^{-1}$}
\newcommand{\chir}{$\chi^2_r$ }

\journalname{Astrophysics and Space Science}

\begin{document}

\title{The first multi-wavelength campaign of AXP 4U~0142+61 from
radio to hard X-rays \thanks{We acknowledge the Director of Gemini for
his DDT and also the Swift team for the ToO. The results in this paper
are based on observations with INTEGRAL.
}}

\author{P.~R. den Hartog \and
        L. Kuiper \and
	W. Hermsen \and
	N. Rea \and
	M. Durant \and
	B.~Stappers \and
	V.~M. Kaspi \and
	R. Dib
}


\institute{P.R. den Hartog \and L. Kuiper \and W. Hermsen \and N. Rea \at 
              SRON Netherlands Institute for Space Research\\
	      Sorbonnelaan 2, 3584 CA Utrecht, The Netherlands\\
              \email{hartog@sron.nl}
           \and
           W. Hermsen \and B. Stappers \at
              Astronomical Institute, University of Amsterdam\\
	      Kruiskade 403, 1098 SJ Amsterdam, The Netherlands
	   \and
	   M. Durant \at
	      Department of Astronomy \& Astrophysics, University of Toronto\\
	      60 Saint George Street, MP 1404D Toronto, ON M5S 3H8, Canada 
	   \and
	   B. Stappers \at
	      ASTRON The Netherlands Foundation for Research in Astronomy\\
	      P.O. Box 2, 7990 AA, Dwingeloo, The Netherlands
	   \and
	   V.M. Kaspi \and R. Dib \at
	      Physics Department, McGill University\\
	      3600 University Street, Montreal, PQ H3A 2T8, Canada
}

\date{Received: \today / Accepted: date}

\maketitle

\begin{abstract}

For the first time a quasi-simultaneous multi-wavelength campaign has
been performed on an Anomalous X-ray Pulsar from the radio to the hard
X-ray band. 4U~0142+61 was an INTEGRAL target for 1 Ms in July
2005. During these observations it was also observed in the X-ray band
with Swift and RXTE, in the optical and NIR with Gemini North and in
the radio with the WSRT. In this paper we present the source-energy
distribution. The spectral results obtained in the individual wave
bands do not connect smoothly; apparently components of different
origin contribute to the total spectrum. Remarkable is that the
INTEGRAL hard X-ray spectrum (power-law index $ 0.79 \pm 0.10$) is now
measured up to an energy of $\sim$230 keV with no indication of a
spectral break. Extrapolation of the INTEGRAL power-law spectrum to
lower energies passes orders of magnitude underneath the NIR and
optical fluxes, as well as the low $\sim$30 $\mu$Jy (2$\sigma$) upper
limit in the radio band.


\end{abstract}

\section{Introduction}
\label{sec:intro}

Anomalous X-ray Pulsars (AXPs) are young rotating isolated neutron
stars \citep[for a review in this volume,
see][]{Kaspi06_london}. Currently there are 8 AXPs known and there are
a few more candidates \citep{Woods06_review}. These objects are called
anomalous, because their X-ray luminosities exceed by far the
available total energy released by rotational energy loss. The energy
output is believed to originate from an immense energy reservoir
stored in a toroidal magnetic field within the Neutron Star.  The
surface magnetic fields, inferred from their periods and period
derivatives, are of the order of $10^{14} - 10^{15}$~G. Therefore,
AXPs are believed to be magnetars, as originally proposed for the Soft
Gamma-ray Repeaters \citep[SGRs, see][]{DT92,TD95,TD96}. Both AXPs and
SGRs are well studied objects in the X-ray band for energies below 10
keV. However, little was known about their persistent emission in the
hard X-ray band ($>$10 keV). In 2004 INTEGRAL discovered hard X-ray
emission from the position of 1E~1841-045 \citep{Molkov04_sagarm}.
\cite{Kuiper04_1841} showed unambiguously that the hard X-rays
originated from the AXP by extracting a pulsed hard X-ray signal from
the source using archival RXTE data. After INTEGRAL discovered hard
X-rays from two other AXP locations, namely from 1RXS~J1708-4009
\citep{Revnivtsev04_gc} and 4U~0142+61 \citep{denHartog04_atel0142},
\cite{Kuiper06_axps} also showed for these and for a fourth AXP
(1E~2259+586) pulsed hard X-ray emission using archival RXTE
data. That means that presently already for 4 of the 8 established
AXPs hard X-ray emission has been detected and this can now be
considered to be a common characteristic, which is not yet understood.

In this paper we focus on the AXP 4U~0142+61. This AXP was discovered
by the Uhuru X-ray observatory in the early seventies
\citep{Giacconi72_uhuru, Forman78_uhuru}.  The spin period of 8.7~s
was found by Israel, Mereghetti \& Stella (1994). They realised that
the X-ray luminosity is too high to be explained by rotational energy
loss.  Like for the other AXPs, there is no proof for a companion, nor
for an active (i.e. accreting) disk that could explain the high X-ray
luminosity of 4U~0142+61. The passive (i.e. non accreting) debris disk
discovered by \cite{Wang06_0142} does not power the X-ray
emission. The X-ray luminosities recently measured with e.g.
XMM-Newton and Chandra are of the order of $10^{35}$ \ecs \citep[2--10
keV, see ][]{Patel03_0142,Goehler05_0142}, assuming a distance of 3.6
kpc \citep{Durant06_distances}. The X-ray spectra (0.5--10 keV) of
AXPs are soft and are commonly fitted with a black-body and a
power-law model. The inclusion of the power-law component is required
to fit excess photons with energies above $\sim$3 keV. For 4U~0142+61,
the best fit parameters are a black-body temperature of $kT \sim$0.4
keV and a power-law photon index of $\Gamma \sim$3.4.

4U~0142+61 was detected by \cite{denHartog06_casa} in hard X-rays up
to 150 keV in 1.6 Ms of INTEGRAL observations \citep[see
also][]{Kuiper06_axps}. The 20--150 keV flux was measured to be $(9.7
\pm 0.9) \times 10^{-11}$ \ecs. The total spectrum could be fitted
with a power-law model with photon index $\Gamma = 0.73 \pm
0.17$. They also revisited the Compton Gamma-Ray Observatory
(COMPTEL) archives \citep[0.75--30 MeV, see][]{Schoenfelder93_comptel}
and determined flux upper limits at the location of the AXP. These
limits put constraints on the extrapolation of the hard X-ray power
law. Assuming that the hard X-ray flux is stable, a spectral break has
to occur in the hard X-ray regime below $\sim$750 keV in order for the
spectrum not to be in conflict with the COMPTEL upper
limits. \cite{denHartog06_casa} and \cite{Kuiper06_axps} were not able
to measure such a break, but a hint was found with a $2.3\sigma$ fit
improvement when a high-energy cutoff was added to the power-law model
by den Hartog et al. (2006a). The indication for the cutoff was found at
an energy of $73 \pm 15$ keV.

A close connection may exist between the production of non-thermal
hard X-rays and radio emission. However, until recently all AXPs were
radio quiet. For AXP 1E~2259+58 the detection of radio emission has
now been claimed by \cite{Malofeev05_2259}, but the observations and
analysis were difficult and confirmation is
required. \cite{Halpern05_1810} discovered transient radio emission
from the transient AXP XTE~J1810-045 which appeared sharply modulated
at the rotation period with peak flux densities $>$1~Jy
\citep{Camilo06_1810}, orders of magnitude brighter than the reported
upper limits for this or any other AXP.  \cite{Gaensler01_snr} have
observed 4U~0142+61 with the VLA (1.4 GHz), but only a $5\sigma$ upper
limit of 0.27 mJy could be extracted from the data.

4U~0142+61 was the first AXP for which an optical counterpart was
discovered (Hulleman, van Kerkwijk \& Kulkarni
2000). \cite{Kern02_0142} found that the optical emission is pulsed
for a considerable fraction. \cite{Hulleman00_0142} showed for the
first time that it is not possible to understand the optical and NIR
measured fluxes with respect to the X-ray fluxes. There is an optical
and NIR excess that can not be explained by the Rayleigh-Jeans tail
from the X-ray black body. Moreover the optical and NIR emissions seem
to be non thermal and exhibit more variability than seen in the X-rays
\citep{Hulleman04_0142, Israel04_opticalaxps, Morii05_0142,
Durant06_0142var}.

We present the first quasi-simultaneous multi-wavelength observation
campaign to study 4U~0142+61 from the radio up to hard X-rays. 

\section{Multi-wavelength campaign}
\label{sec:mmc}

\begin{table}[b]
\caption{Multi-wavelength campaign measurements}
\centering
\label{tab:mobs}      
\begin{tabular}{lll}
\hline\noalign{\smallskip}
Obs & Time span & Exposure  \\[3pt]
\tableheadseprule\noalign{\smallskip}
WSRT & July 2, 2005& 12 h\\
Gemini $K_s$& July 26, 2005& 1125 s\\
Gemini $r^\prime$& July 13, 2005& 2400 s\\
Swift & July 11 -- 12, 2005 & 7400 s\\
INTEGRAL & June 29 -- July 17, 2005 & 868 ks \\
\noalign{\smallskip}\hline
\end{tabular}
\end{table}

For June--July 2005, 1 Ms INTEGRAL (20--300 keV) dedicated 4U~0142+61
observations were scheduled. With these observations, we tried to get,
nearly simultaneously, as much wavelength-band coverage of 4U~0142+61
as possible. An approved 12 hour radio observation with the WSRT (21
cm) was rescheduled to overlap with the INTEGRAL observations.  A
regular RXTE (2--250 keV) monitoring observation also fell in the
INTEGRAL time line. For an X-ray imaging observation a Target of
Opportunity (ToO) was granted with Swift (0.2--10 keV).
Finally, optical and NIR observations were requested and approved in
the Directors' Discretionary Time (DDT) on Gemini North. During
two nights 4U~0142+61 was imaged in the $K_s$ and $r^\prime$
bands. Unfortunately it was not possible to schedule the $K_s$ and the
INTEGRAL observations contemporaneous (see Table~\ref{tab:mobs}).

\subsection{Hard X-rays: INTEGRAL}
\label{sec:igr}

The INTErnational Gamma-Ray Astrophysics Laboratory
\citep[INTEGRAL;][]{Winkler03_igr} is ESA's currently operational hard
X-ray/soft gamma-ray space telescope. For the study of AXPs the
low-energy detector of IBIS, called ISGRI \citep[20--300
keV][]{Lebrun03_isgri}, has proven itself to be of great importance.
The serendipitous discovery of AXPs in the INTEGRAL energy band was a
result of the combination of long exposure times and the $29^\circ$
wide FOV of IBIS.

For this work we have analysed 1 Ms of observations of 4U~0142+61
performed between June 29 and July 17, 2005. The data were screened for
solar flares and erratic count rates resulting from passes through the
Earth's radiation belts. After screening the net exposure was 868 ks
(Table~\ref{tab:mobs}). The observations consist of 265 pointings
(Science Windows, ScWs) which can last up to one hour. The ScWs have
been analysed separately with the official INTEGRAL software OSA~5.1
\citep[see][for IBIS-ISGRI scientific data analysis]{Goldwurm03_osa}
in 20 energy bands between 20 keV and 300 keV with exponential
binning. These analyses result in sky images for every ScW in 20
energy bands. The spectrum was built up by averaging the count rates
from each ScW, weighted by the variance. For the conversion into flux
values, the spectrum was normalized to the known total Crab spectrum
(nebula and pulsar) using a curved power law as determined by
\cite{Kuiper06_axps}.

4U~0142+61 is detected up to 230 keV with a 3.0$\sigma$
significance in the 150--230 keV energy band.  The total spectrum
shown in Fig.~\ref{fig:sed} was fitted with a power-law model
resulting in a photon index $\Gamma = 0.79 \pm 0.10$.  The quality of
the fit is good with a reduced chi square \chir = 1.12 for 16 degrees
of freedom. The 20--230 keV flux is $(17.0 \pm 1.4) \times 10^{-11}$
\ecs.

Our new total spectrum shows that this AXP is now detected at even
higher energies than reported earlier, without an indication for a
spectral break.  The hint for a break found by \cite{denHartog06_casa}
is not confirmed in this data set.

\subsection{Soft X-rays}
\subsubsection{Swift}
\label{sec:swift}

The Swift-XRT \citep[0.2--10 keV; ][]{Burrows05_xrt} observation
was performed on July 11--12, lasting 8500 s. Of this observation 7400
s of data were taken in the Photon-Counting mode and were analysed
with the FTOOLS {\tt xrtpipeline}, version build-14 under HEADAS 6.0
(Hill et al. 2005).  Photons were extracted from an annular region (3
pixels inner radius, 30 pixels outer radius) in order to avoid pile-up
contamination. Background spectra were taken from close-by source-free
regions.

As mentioned in Sect.~\ref{sec:intro}, AXP spectra in the X-ray domain
($<10$ keV) can usually be fitted satisfactorily with a black-body and
a power-law model. When we use this canonical model the fit results
are: $N_{\rm H} = (1.01 \pm 0.10) \times 10^{22} {\rm cm^2}$; $kT =
(0.400 \pm 0.012)$ keV; $\Gamma = 2.7 \pm 0.3$; \chir = 0.97 (dof =
573).  These parameters yield an unabsorbed flux of $(3.9 \pm 0.2)
\times 10^{-10} \,$ \ecs\ in the 0.7--6.0 keV band, or $(7.8 \pm 0.5)
\times 10^{-11} \,$ \ecs\ in the more commonly used 2--10 keV band.
Such a model fits the measured X-ray spectrum well, but systematically
overestimates the flux at the lower X-ray energies. Furthermore it
seems meaningless for extrapolation to the NIR window. In particular,
the soft power-law component dominates the black-body component for
energies less than $\sim$1.5 keV, meaning that also the $N_{\rm H}$
estimate is affected and estimated too high.

Alternatively, we used a double black-body model to fit the measured
spectrum, yielding an acceptable fit (\chir = 0.98; dof = 573) with an
excellent fit at the lower X-ray energies, but underestimating the
higher X-ray energies ($>4.5$ keV).  The two temperatures are $0.38
\pm 0.02$ keV and $0.78 \pm 0.10$ keV and the $N_{\rm H}$ is $(0.61
\pm 0.03) \times 10^{22} {\rm cm^{-2}}$. The unabsorbed flux is $(2.11
\pm 0.06) \times 10^{-10}$ \ecs\ in the 0.7--6.0 keV band and $(6.94
\pm 0.29) \times 10^{-11}$ \ecs\ in the 2--10 keV band.  Arguably,
this model is again canonical, but it is more appropriate for
extrapolation to the NIR, which is shown in Fig.~\ref{fig:sed}.

\subsubsection{Rossi-XTE}
\label{sec:rxte}

4U~0142+61 monitoring data of the PCA \citep[2--60
keV;][]{Jahoda96_pca} on board the Rossi X-ray Timing Explorer
\citep[RXTE, ][]{Bradt93_rxte} were used to create a phase-coherent
timing solution valid during the multi-wavelength campaign
\citep[see][for a detailed description and use of the AXP-monitoring
campaign]{Gavriil02_rxtemonitoring}. This ephemeris is essential for
INTEGRAL timing analysis \citep[c.f., ][]{denHartog06_0142}. In this
work it is used for the WSRT observation in order to reduce the number
of trials in the search for a (possible) weak pulsed radio
signal. Data from RXTE-observation IDs 90076 and 91070 were analysed
with the pulsar-timing software package
TEMPO\footnote{http://pulsar.princeton.edu/tempo}.  The resulting
ephemeris is valid between MJD 53251 and MJD 53619, with the following
characteristics: Epoch MJD 53420.0, $\nu = 0.1150929855(7)$ Hz,
$\dot{\nu} = -2.639(8) \times 10^{-14}\, {\rm Hz\,s^{-1}}$ and $
\ddot{\nu} = 3(2) \times 10^{-23}\, {\rm Hz\,s^{-2}}$.

\subsection{Optical \& NIR: Gemini}
\label{sec:nir}

The field of 4U~0142+61 was imaged in the $K_s$ band with the Near
Infrared Imager \citep[NIRI;][]{Hodapp03_gemini} on Gemini
North, Hawaii, in the night of July 26th, 2005. NIRI has
standard broad-band and narrow-band filters covering 1--5$\mu m$.

The final image was created by subtracting dark frames from each
science frame, dividing by a flat field derived from the images
themselves, and then aligning and stacking all the images. The
photometry was measured using the PSF-fitting package DAOPhot
\citep{Stetson87_daophot}, and calibrated against the photometry
provided for several field stars in \cite{Hulleman04_0142}.  The $K_s$
magnitude was found to be 19.96(10).

Optical $r^\prime$-band images were obtained on the night of July
13th, 2005 using the Gemini-North Multi-Object Spectrograph
\citep[GMOS-N;][]{Hook04_gemini}. We subtracted the bias and divided
by screen flats, before stacking and photometering the images. For the
calibration, the photometry listed in Hulleman, van Kerkwijk \&
Kulkarni (2004) was used,
interpolating between the R and V-bands using the relationship in
\cite{Smith02_conversion}.  We find for 4U 0142+61 $r'=25.42(6)$,
where the statistical uncertainty in the measurement and the
uncertainty in the calibration of the photometry zero-point are
similar.

\subsection{Radio: WSRT}
\label{sec:radio}

Using the Westerbork Synthesis Radio Telescope (WSRT) we have
searched for both pulsed and unpulsed radio emission. An observation
of 12 hour duration was carried out at a frequency of 1380 MHz ($\sim$
21 cm) with a bandwidth of 80 MHz. Using the synthesis data a map was
made using standard routines in the
MIRIAD\footnote{http://www.atnf.csiro.au/computing/software/miriad/}
package. In the resulting map (rms $\sim$30 $\mu$Jy) no source was
detected at the location of 4U 0142+61 leading to a $3\sigma$ upper
limit on its flux of $\sim$90 $\mu$Jy.  Simultaneously we also summed
the signals from all 14 dishes of the WSRT coherently to form a
so-called tied array. The data were then sent to the Pulsar
Machine PuMa \citep{Voute02_puma} which formed a digital filter bank
with 512 channels and a sampling time of 409.6~$\mu$s. As the
dispersion measure in the direction of the source was unknown, we
tried many trial dispersion measures and then folded each one with the
RXTE ephemeris (see Sect.~\ref{sec:rxte}). Each resultant profile was
then inspected to determine if there was a significant detection. We
also performed a standard pulsar search analysis on the full data
set. Neither the folding nor the search revealed any significant
detection of radio pulsations.  A $5\sigma$ upper limit was determined
at $77 \times \sqrt(d/1-d) \mu{\rm Jy}$ where $d$ is the duty cycle of
the pulsar. Using $d = 0.5$, a typical value in the X-ray regime, the
$3\sigma$ upper limit is $\sim$46 $\mu{\rm Jy}$. It has to be noted
that for this analysis the whole observation was used. A finer
analysis like performed by \cite{Halpern05_1810} and
\cite{Camilo06_1810} who discovered the transient AXP XTE~J1810-197 as
a bright transient radio source in smaller intervals is still
ongoing. However 4U~0142+61 has not exhibited the same sort of
transient behaviour in X-rays like XTE~J1810-197, specially it has not
shown any large outburst, and therefore the radio characteristics of
these sources might be very different.

\section{Multi-wavelength Source-Energy Distribution}
\label{sec:sed}

\begin{figure*}
\centering
\begin{minipage}[c]{0.65\textwidth}
\psfig{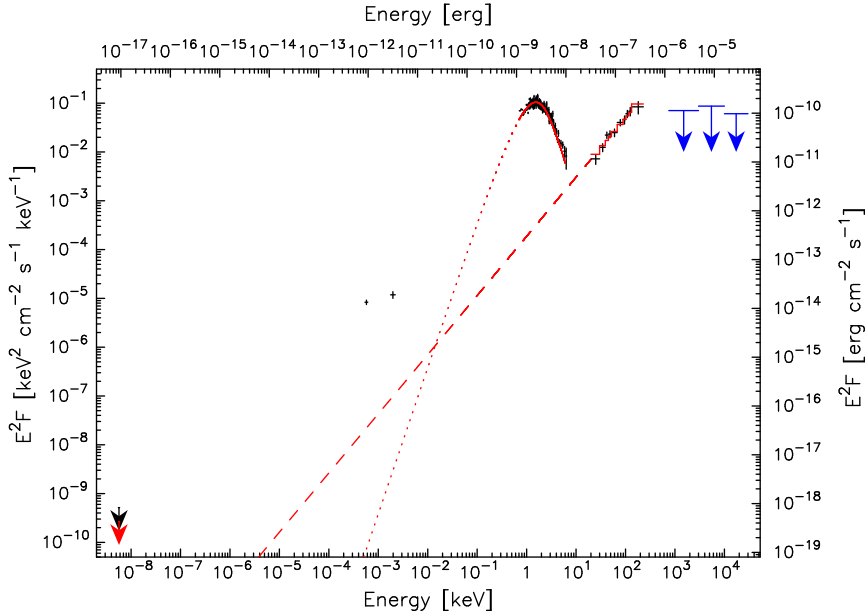}
\end{minipage}%
\caption{Source-energy distribution of 4U~0142+61 from radio up to
gamma-ray energies. For observation details, see
Sect.~\ref{sec:sed}. The WSRT radio limits are shown for the continuum
emission (top arrow) and the pulsed emission (lower
arrow). Extrapolations to lower energies are shown for the Swift-XRT
double black-body fit (dotted line) and for the INTEGRAL power-law fit
(dashed line). It can be seen that neither the soft X-ray nor the hard
X-ray spectral fits extrapolate to the Optical and NIR fluxes. An
optical-NIR excess orders of magnitude above these extrapolations is
evident. The COMPTEL upper limits at MeV energies do not belong to the
multi-wavelength campaign.}
\label{fig:sed}      
\end{figure*}

In Fig.~\ref{fig:sed} the quasi-simultaneous multi-wavelength
Source-Energy Distribution (SED) is presented covering roughly 10
orders of magnitude in photon energy. In the lower left corner the
WSRT (1380 MHz) continuum (upper arrow) and timing (lower arrow)
$2\sigma$ upper limits are shown. The Gemini $K_s$ (2.15 $\mu$m)
and $r^\prime$ (0.6 $\mu$m) data points are dereddened assuming $A_V =
3.6$ as determined by \cite{Durant06_extinction}. The Swift-XRT
spectrum between 0.7 keV and 6 keV is corrected for absorption with
$N_{\rm H} = 0.61 \times 10^{22} {\rm cm^{-2}}$. Also shown is the
corresponding double black-body fit (solid line), extrapolated towards
lower energies (dotted line). The INTEGRAL flux values between 20 keV
and 230 keV show the hard X-ray spectrum for this AXP.  The power-law
fit (solid line) is also extrapolated to lower energies (dashed line).

Significant flux variability has been reported in the
optical, NIR and soft X-ray bands \citep{Durant06_0142var},
therefore in this quasi-simultaneous spectrum we can investigate
better how the fluxes in the different bands relate to each other.

\section{Discussion}
\label{sec:discussion}

The results for the different wave-band measurements render separately
no surprises.  The Swift-XRT spectrum shows a typical soft AXP
spectrum and NIR and optical magnitudes are around the earlier
reported values. The INTEGRAL spectrum is in agreement with previously
found results, however, thanks to the high effective exposure in this
dedicated observation the maximum energy up to which 4U~0142+61 could
be detected is now higher, namely 230 keV.  The spectrum, described
with a power-law model with photon index $\Gamma = 0.79 \pm 0.10$ and
luminosity $ 8.7 \times 10^{34} \,{\rm erg\, s^{-1}}$ (20--100 keV, $d$
= 3.6 kpc) can be compared with the spectral results reported earlier
by \cite{denHartog06_casa} using an independent data set: $\Gamma =
0.73 \pm 0.17$ and luminosity $8.5 \times 10^{34}\, {\rm erg\, s^{-1}}$
(20--100 keV, $d$ = 3.6 kpc).  \cite{Kuiper06_axps} derived $\Gamma =
1.05 \pm 0.11$ and luminosity $8.1 \times 10^{34} \,{\rm erg\, s^{-1}}$
(20--100 keV, $d$ = 3.6 kpc).  All these findings are within errors in
agreement. Therefore there is no evidence for long-term time
variability at hard X-ray energies yet.

\cite{denHartog06_casa} published $2\sigma$ flux upper limits for 4U
0142+61 in the 0.75--30 MeV window analysing COMPTEL data collected
over the years 1991--2000. These observations are obviously not
contemporaneous to the multi-wavelengths campaign reported here, but
the apparent stability at hard X-ray energies seems to justify a
direct comparison of the COMPTEL upper limits with the hard X-ray
spectrum measured with INTEGRAL.  Thus assuming that the hard X-ray
emission is stable, Fig.~\ref{fig:sed} clearly shows that the hard
X-ray / soft gamma-ray spectrum of 4U0142+61 has to break between
$\sim$ 200 keV and 750 keV in order not to be in conflict with the
COMPTEL upper limits. If we were to assume that by chance 4U0142+61
was in a low state at MeV energies during the long COMPTEL
observations, it would be remarkable that also for the other AXPs
detected similarly by INTEGRAL (1E 1841-045 and 1RXS J1708-4009) no
signal was found at MeV energies during the many observations that
they were in the COMPTEL field-of-view spread over 10 years
\citep{Kuiper06_axps}.  We conclude that a drastic break is required.

The Swift-XRT and INTEGRAL spectra can not be fitted
simultaneously with a two-component model consisting of i.e. a black
body plus only one power law component. The excess of high-energy
photons (w.r.t a single black-body spectrum) in the Swift
spectrum can not be accounted for by the upcoming hard X-ray
spectrum. This is evident in Fig.~\ref{fig:sed}. The extrapolation of
the hard X-ray spectral fit towards lower energies is already five
times lower at 6 keV than measured with Swift.

The relation of the optical and NIR flux values to the soft and hard
X-ray spectra is still an enigma. It was noted 
earlier \citep[e.g., ][]{Hulleman04_0142} that extrapolations of fits
to the soft X-ray spectra to the optical and NIR bands are not
consistent with the measured variable flux values. This is more than
evident in Fig.~\ref{fig:sed}, where we show the extrapolation of the
double black-body fit, which reaches optical and NIR fluxes $\sim$4--5
orders of magnitude lower than the measured values.  The very hard
X-ray spectra above 10 keV, originally interpreted as being of
non-thermal origin, led to suggestions of a common origin as the
likely non-thermal optical and NIR emissions. However,
Fig.~\ref{fig:sed} unambiguously shows that also extrapolation of the
power-law spectral shape measured at hard X-ray energies passes 2--3
orders of magnitude underneath the optical and NIR data points. The
hard X-ray, soft X-ray, optical and NIR components seem to have
different origins.

The discovery of the hard X-ray spectra of AXPs, stimulated new
attempts to search for non-thermal radio emission possibly originating
from the same sites and/or production mechanisms (e.g. synchrotron
radiation) in the magnetar's magnetosphere. The WSRT upper limits at
1380 MHz are however not constraining in this respect. The INTEGRAL
power-law fit extrapolates also orders of magnitudes below the radio
upper limits.

The physical interpretation of the hard X-ray spectrum measured by
INTEGRAL is not evident. Assuming a distance of 3.6 kpc as recently
determined by Durant \& van Kerkwijk (2006a), the 20--230 keV flux
(Sect.~\ref{sec:igr}) translates to a luminosity of $2.6 \times
10^{35}\, {\rm erg\, s^{-1}}$, which exceeds the maximum luminosity
available from rotational energy loss\footnote{$L_{\rm RE} = 1.3
\times 10^{32}\, \rm{erg\, s^{-1}}$ assuming a neutron star with
$R=10$ km, $M=1.4 \rm{M_\odot}$} with a factor of $\sim$2000.
Moreover, the hard X-ray luminosity is comparable with the high soft
X-ray luminosity ($1.1 \times 10^{35}\, {\rm erg\, s^{-1}}$, 2--10
keV) as measured with Swift (see Sect.~\ref{sec:swift}).

Over the last two years theoretical attempts have been made trying to
explain the new hard X-ray emission above 10 keV from AXPs.  A
particularly interesting interpretation of the high-energy emission is
the application of a magnetar-corona model by Thompson \& Beloborodov
(2005) and Beloborodov \& Thompson (2006a,b in this volume). In the
last papers, a theoretical model is developed explaining the formation
of a hot corona above the surface of a magnetar, so far the only model
attempting to explain the multi-wavelength emission. In essence, the
twisted magnetosphere acts as an accelerator that converts the
toroidal-field energy to particle kinetic energy. They show
numerically that the corona self organises quickly into a quasi-steady
state, with a voltage of $\sim$1 GeV along the magnetic field
lines. Pair production occurs at a rate just enough to feed the
electric current. Interestingly, the heating rate of the corona is
$\sim$$10^{36}$--$10^{37} {\rm erg\, s^{-1}}$, in agreement with the
observed persistent, high-energy output of magnetars. Furthermore,
they deduce that the static twist will decay on a timescale of 1--10
years, setting the scale for time variability to look for in the
high-energy emission. The transition layer between the atmosphere and
the corona is the likely source of the observed $\sim$100 keV
(e.g. bremsstrahlung) emission from magnetars. Finally, it is worth
mentioning that the corona emits curvature radiation which can also
supply the observed IR-optical luminosity. Of particular concern in
the hot coronae model is the reported timescale of 1--10 years. For 4U
0142+61 we do not see evidence for strong variability in the hard
X-ray emission over the first years of INTEGRAL observations, nor over
the longer period of RXTE monitoring observations. Future work will
concentrate on this aspect, as well on detailed timing studies,
including phase resolved spectroscopy. Obviously, as much INTEGRAL
data as possible will be collected on 4U~0142+61 and the other hard
X-ray emitting AXPs to search for the break energy in the total
spectrum, a key parameter in the discrimination between the different
proposed models.  More detailed analysis of this campaign and
additional INTEGRAL observations will be presented in forthcoming
papers by \cite{Rea06_0142} and \cite{denHartog06_0142}, respectively

\begin{acknowledgements}
We acknowledge J. Vink for the use of private INTEGRAL Cas~A data.
\end{acknowledgements}



\end{document}